\documentclass{aastex6}

\fullcollaborationName{The Friends of AASTeX Collaboration}

\begin{document}

\title{Size Distribution of Small Hilda Asteroids}

\author{
Tsuyoshi Terai\altaffilmark{1}
and
Fumi Yoshida\altaffilmark{2,3}
}
\email{tsuyoshi.terai@nao.ac.jp}
\footnote{Based on data collected at Subaru Telescope, which is operated by the National 
Astronomical Observatory of Japan.}

\altaffiltext{1}{Subaru Telescope, National Astronomical Observatory of Japan, National Institutes 
of Natural Sciences (NINS), 650 North A`ohoku Place, Hilo, HI 96720, USA}
\altaffiltext{2}{Planetary Exploration Research Center, Chiba Institute of Technology, 
2-17-1 Tsudanuma, Narashino, Chiba 275-0016, Japan}
\altaffiltext{3}{Department of Planetology, Graduate School of Science, Kobe University, Kobe, 
657-8501, Japan}

\begin{abstract}

We present the size distribution for Hilda asteroid group using optical survey data obtained by the 
8.2~m Subaru Telescope with the Hyper Suprime-Cam.
Our unbiased sample consists of 91 Hilda asteroids (Hildas) down to 1~km in diameter.
We found that the Hildas' size distribution can be approximated by a single-slope power law in 
the $\sim$1$-$10~km diameter range with the best-fit power-law slope of $\alpha$~=~0.38~$\pm$~0.02 
in the differential absolute magnitude distribution.
Direct comparing the size distribution of Hildas with that of the Jupiter Trojans measured 
from the same dataset \citep{YT17} indicates that the two size distributions are well similar to 
each other within a diameter of $\sim$10~km, while these shapes are distinguishable from that of 
main-belt asteroids.
The results suggest that Hildas and Jupiter Trojans share a common origin and have a different 
formation environment from main-belt asteroids.
The total number of the Hilda population larger than 2~km in diameter is estimated to be 
$\sim$1~$\times$~$10^4$ based on the size distribution, which is less than that of the Jupiter 
Trojan population by a factor of about five.

\end{abstract}


\section{Introduction} \label{sec01}

The Hilda group is an asteroid population located in the 3:2 mean motion resonance with Jupiter, 
near $\sim$4.0~au from the Sun.
Although its current location is just an extension of the main asteroid belt, Hilda asteroids
(Hildas) are quite different from main-belt asteroids (MBAs) from the composition's point of 
view. 
The Hilda group is populated by low albedo asteroids such as the C, P, and D types \citep{Gr89} 
and no S-type asteroids which is common among MBAs \citep{DC13}. 
The taxonomic distribution of Hildas is similar with that of Jupiter Trojans (JTs) rather than that 
of MBAs \citep{DC14}. 
Since the recent dynamical models for the solar system formation \citep[e.g.,][]{Mr05,Lv09,Ns13} 
claimed that the present JTs were captured into the Jupiter Trojan's orbits from the transplanetary 
planetesimal disk during the planet migration, the similarity of the taxonomic distribution between 
Hildas and JTs also attracts our interest from the point of dynamical evolution of our solar system.

JTs have been well studied than Hildas so far and are known to possess a bimodality in the spectral
distribution. 
\citet{Sz07} found that their color distribution is significantly different from that of MBAs and 
appears bimodal using the Sloan Digital Sky Survey (SDSS) Moving Object Catalog (MOC).
\citet{Ro08} reported a clear bimodality in the distribution of visible spectral slopes among 
JTs, two thirds of which consist of reddish objects corresponding to D-type asteroids and the 
remaining bodies show less reddish colors compatible with the P-type.
The bimodal distribution was also confirmed in the near-infrared spectra \citep{Em11}.

The bimodality in spectral distribution indicates that JTs consist of two sub-populations, which 
provides useful insight for understanding their origin.
\citet{WB16} proposed an interesting hypothesis related to this bimodality that location-dependent 
volatile loss from planetesimals as a progenitor population of JTs and trans-neptunian objects 
(TNOs) through sublimation could have divided them into two color sub-populations. 
They focused on the surface abundance of H$_2$S ice as the primary factor in creating JT's color 
bimodality.
Objects that formed beyond the H$_2$S sublimation line (around 20~au) retained H$_2$S ice on their 
surfaces which contributes to more significant reddening of the spectra through irradiation compared
to the other sub-population of objects with H$_2$S depletion located closer than this line.
The two sub-populations were scattered throughout the middle and outer solar system during the period 
of dynamical instability between giant planets, and their mixture in the JT region brought the 
bimodal color distribution.

The Hilda population is possibly originated from these two sub-populations of primordial 
planetesimals because they also exhibit bimodality in spectra \citep{GB08,WB17a,Dr18}.
\citet{Wn17} showed that the corresponding color sub-populations (less-red and red) within JTs and 
Hildas have similar spectral shapes in the near-infrared range, suggesting a common progenitor 
population between the two groups.
\citet{TE12} found that the spectra of P-type asteroids in the Hilda and Cybele groups are 
characterized by a ``rounded" 3-$\mu$m absorption feature, which resemble those of less-red JTs 
\citep{Br16}.
This absorption could be due to the presence of H$_2$O frost on the surfaces implying that these 
objects did not experience aqueous alteration \citep{TE12}, but could also be due to N-H stretch 
features implying that these objects formed beyond the giant planet region where NH$_3$ would 
have been stable \citep{Br16}.
According to numerical simulations regarding the orbital evolution and chaotic capture of 
trans-Jovian objects due to the gravitational instability, a significant number of them are trapped 
into the JT, Hilda, and the outer MBA regions \citep[e.g.,][]{LH10}.
These facts support the common origin of Hildas and JTs from trans-Jovian objects.

In this paper, we compare the size distributions of Hildas and JTs to show further possible 
similarity between the two populations.
The size distribution of an asteroid population is initially determined by a characteristic of the
accretion process and then is altered by cascade fragmentation through subsequent collisional 
evolution, while it is usually insensitive to dynamical evolution and gravitational disturbance.
Therefore, asteroid groups sharing the same formation region should have similar size distributions.

Several projects such as Spitzer \citep{RW11}, WISE \citep{Gr12a}, and SDSS \citep{WB17a}, have done 
surveys for Hildas and measured their size distribution.
\citet{RW11} reported the size distribution with a break at diameter $D$~$\sim$~12~km, but 
\citet{Gr12a} showed that a single-slope power law is a much better fit to Hildas' size 
distribution in $D$~$>$~5~km.
\citet{WB17a} found that the two sub-populations, less-red and red Hildas, exhibit distinct size 
distributions from each other in $D$~$\gtrsim$~5~km.
Such discrepancy is also seen in JTs, i.e., the size distribution of red JTs is shallower than 
that of less-red JTs \citep{WB15}.
Furthermore, the dynamically excited TNOs with high inclination and eccentricity, so-called hot 
TNOs, have been confirmed to exhibit a bimodal color distribution \citep{FB12} and share the 
comparable size distribution to JTs over the common observable size range \citep{Fr14}.
The size distributions of the two sub-populations of hot TNOs are also nearly identical in shape
\citep{WB17b}. 
Considering these facts, Hildas are believed to originate from a common progenitor population with 
JTs and hot TNOs.
However, it is still uncertain whether the size distribution of Hildas is equivalent to that of
JTs, particularly in the size range dominated by their collisional evolution.
 
Recently, \citet{YT17} performed the L4 JT swarm survey using the 8.2-m Subaru Telescope, revealing 
the characteristic of JTs' size distribution with a completeness diameter limit of 2~km that 
exhibits a single-slope power law within a diameter of $\sim$10~km and becomes steeper beyond this 
range.
As a next step, the size distribution of Hildas should be determined in such a size range.
However, the sample sizes of the previous works for Hildas are insufficient for comparison of the 
size distribution with JTs due to the low sensitivities relative to that of \citet{YT17}.
Therefore, we investigate the Hildas' size distribution in the same dataset as \citet{YT17}, which 
allows us to detect Hildas much smaller than the previous surveys.
This study is capable of highly accurate evaluation of similarities and differences in the size 
distribution between the Hilda and JT populations.

Additionally, we note the shape characteristics of size distribution between Hildas and MBAs.
Although, as mentioned above, the taxonomic distribution of Hildas is distinct from that of MBAs,
P-type asteroids occupy a certain fraction of the outer MBAs next to C-type asteroids \citep{DC13}.
According to the so-called ``Grand Tack" model, where the inward-then-outward migration of Jupiter 
and Saturn implanted asteroids from between and beyond the giant planets into the inner region
\citep{Wl11,Wl12}, it cannot be ruled out that a major portion of the outer MBAs and Hildas are 
originated from the common regions.
Furthermore, their reflective spectra are possible to be altered by surface processes such as 
heating/aqueous alteration \citep[e.g.,][]{Jn90,Fr14} and space weathering \citep{Ln18}.
Therefore, we should compare not only the spectral distributions, but also the size distributions 
among MBAs, Hildas, and JTs for discussing their origins.
It allows us to deepen our understanding of the dynamical evolution processes of planetesimals in 
the inner/middle solar system during or following giant planet migration.

\section{Observation and Data Analysis} \label{sec02}

Our observation was conducted with the Hyper Suprime-Cam (HSC) mounted on the 8.2-m Subaru Telescope 
on March 30, 2015 (UT).
The HSC consists of 116 2k~$\times$~4k Hamamatsu fully depleted CCDs, which covers a 1.5$\arcdeg$ 
diameter field-of-view (FOV) with a pixel scale of 0$\farcs$17 \citep{My12}. 
We surveyed $\sim$26~deg$^{2}$ of sky with 240-sec exposures in the $r$-band. 
The observed area was centered at RA~=~12$^{\rm h}$33$^{\rm m}$ and Dec~=~$-$3$\arcdeg$00$\arcmin$,
located within 4$\arcdeg$ from the ecliptic and within 5$\arcdeg$ of the opposition, 
corresponding to $\sim$10$\arcdeg$$-$20$\arcdeg$ from the Jupiter L4 point. 
Each field was visited three times with time intervals of $\sim$18~min and 36$-$56~min.
The typical seeing size was 0$\farcs$6$-$1$\farcs$0, but partially larger than 1$\farcs$2.

The data were processed with \textit{hscPipe} \citep[version~3.8.5;][]{Bs17}, the image analysis 
pipeline for HSC data based on the Large Synoptic Survey Telescope (LSST) pipeline software 
\citep{Iv08, Ax10}.
We extracted moving object candidates from the source catalogs created by \textit{hscPipe} with the 
westward sky motion ranging from $\sim$~16~arcsec~hr$^{-1}$ to $\sim$~28~arcsec~hr$^{-1}$ targeting 
Hildas and JTs.
All of the candidate sources were visually inspected and 761 asteroids were identified.
Then, we divided these objects into the two groups, Hildas and JTs, based on whether the 
ecliptic-longitudinal motion is faster than that of a circular orbit with semi-major axis of 4.5~au 
or not; the faster are Hildas, and the others are JTs.
As a result, 130 asteroids in the sample were classified as Hildas.

\begin{figure}
\figurenum{1}
\epsscale{0.50}
\plotone{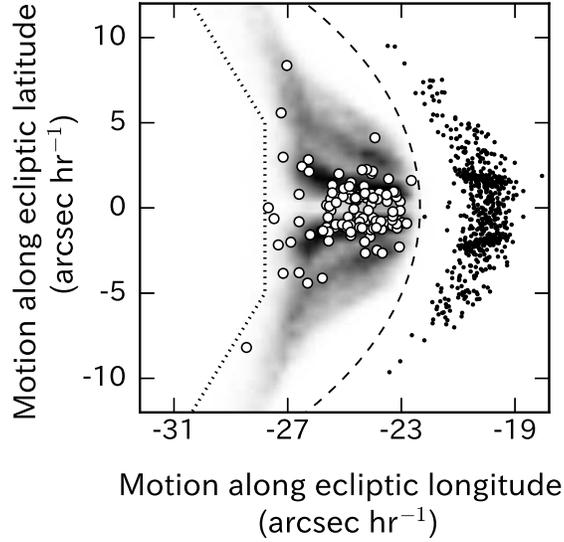}
\caption{
Motions along ecliptic longitude/latitude of the detected Hildas (open circles) and JTs (dots) in 
our survey. 
The background density map is the two-dimensional motion histogram of synthetic Hildas generated 
based on the known orbital distributions by the Monte Carlo method.
The dotted line shows the upper motion limit of our moving object search. 
The dashed line shows the boundary between Hildas and JTs that we defined.
\label{fig01}
}
\end{figure}

Figure~\ref{fig01} shows the detected Hildas and JTs with the separation boundary.
The background density map represents the two-dimensional motion histogram of synthetic Hildas
randomly generated under the probability distribution based on the orbits of the known Hildas.
The motion distribution of our own Hilda sample well agrees with the reproduced plot.

However, the Hilda sample selected by only the sky motion in a short observation arc may contain
a contamination from other asteroid groups, mainly MBAs.
We estimated the fraction of MBAs intruding into the motion range for selecting Hildas using 
synthetic orbits generated in the same manner as above.
The number density of detectable MBAs, which are defined as objects larger than 1~km in diameter,
is given as 100~deg$^{-2}$ based on the previous Subaru/Suprime-Cam ecliptic surveys
\citep{Ys03, YN07}.
The expected numbers of the intruders are about 0.06, 0.12, and 0.23 from the inner (2.0$-$2.6~au),
middle (2.6$-$3.0~au), and outer belts (3.0$-$3.5~au), respectively, in each FOV.
This result corresponds to the total contamination rate of no more than $\sim$5~percent.

Aperture photometry for the detected objects was performed by the same method as \citet{YT17}.
The measured flux was converted into apparent AB magnitude using the photometric zero-points 
estimated by referring to the Panoramic Survey Telescope and Rapid Response System (Pan-STARRS) 
catalog \citep{Sc12, Tn12, Mg13} in the \textit{hscPipe} processing.
The apparent magnitude $m$ was converted into absolute magnitude $H$ through an equation of 
$H = m - 5 \log ( r \Delta ) - P(\theta)$, where $r$ and $\Delta$ are the heliocentric and 
geocentric distances in au, respectively.
$P(\theta)$ is the $H$-$G$ photometric phase function \citep{Bw89} at a solar phase angle
$\theta$ assuming a slope parameter ($G$) of 0.15. 

\begin{figure}
\figurenum{2}
\epsscale{0.70}
\plotone{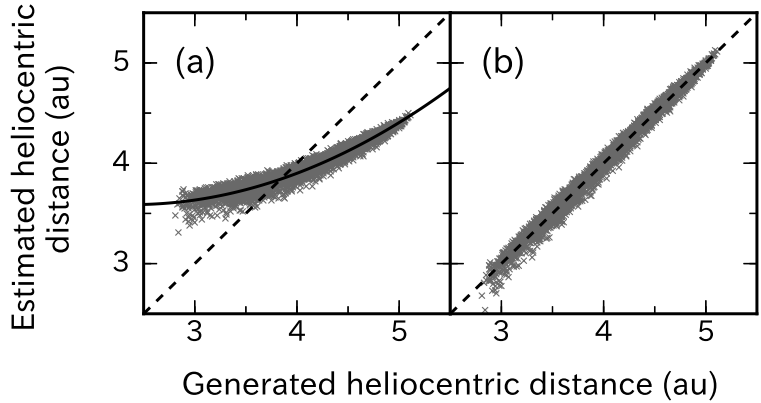}
\caption{
(a) Correlation plot of heliocentric distances between generated synthetic Hilda orbits by
Monte Carlo technique and estimated orbits with circular orbit assumption (see text). 
The solid line shows the best-fit quadratic function. 
(b) Same as panel (a), but the ordinate represents the corrected heliocentric distance.
\label{fig02}
}
\end{figure}

The heliocentric distance of each object was estimated from the measured sky motion assuming
a circular orbit using the expressions presented in \citet{Tr13}. 
To evaluate the estimation accuracy, we conducted a Monte Carlo simulation of the orbit 
calculation by generating 50,000 synthetic Hilda orbits.
Each of the synthetic orbits with a heliocentric distance $r_{\rm act}$ was converted into a virtual 
sky motion based on the identical observation configuration as our survey, and then we estimated the 
heliocentric distance ($r_{\rm est}$) from this sky motion with circular orbit assumption.
Accuracy of the orbital estimation was evaluated by the discrepancy between $r_{\rm act}$ and 
$r_{\rm est}$.
The result showed a significant systematic deviation between them as seen in Figure~\ref{fig02}(a).
This relation can be approximated by a quadratic expression of
\begin{equation}
r_{\rm est} \, = \, 0.120 \, r_{\rm act}^2 \, - \, 0.570 \, r_{\rm act} \, + \, 4.267,
\label{eq01}
\end{equation}
which was derived from a least square fitting.
After removing the systematic error through this equation as seen in Figure~\ref{fig02}(b),
the root-mean-square of ($r_{\rm est} - r_{\rm act}$) is 0.19~au, which causes uncertainties in the
absolute magnitude and body diameter of 0.24~mag and 11\%, respectively.
We applied this correction to the observed Hildas and estimated their heliocentric/geocentric 
distances.

The absolute magnitude is converted into body diameter $D$ by
\begin{equation}
\log D = 0.2 m_\sun + \log ( 2 r_\earth ) - 0.5 \log p - 0.2 H,
\label{eq02}
\end{equation}
where $m_\sun$ is the apparent $r$-band magnitude of the Sun \citep[$-$26.91~mag;][]{Fk11} and
$r_\earth$ is the heliocentric distance of the Earth (i.e., 1~au) in the same unit as $D$.
In this paper, the geometric albedo $p$ is assumed to be a constant of 0.055, which is the mean 
value of Hildas from the NEOWISE measurements presented by \citet{Gr12a}.
Note that \citet{Gr12a} found no significant size-albedo dependency for Hildas.

\section{Results} \label{sec03}

\subsection{Sample selection} \label{sec03-1}

\begin{figure}
\figurenum{3}
\epsscale{0.50}
\plotone{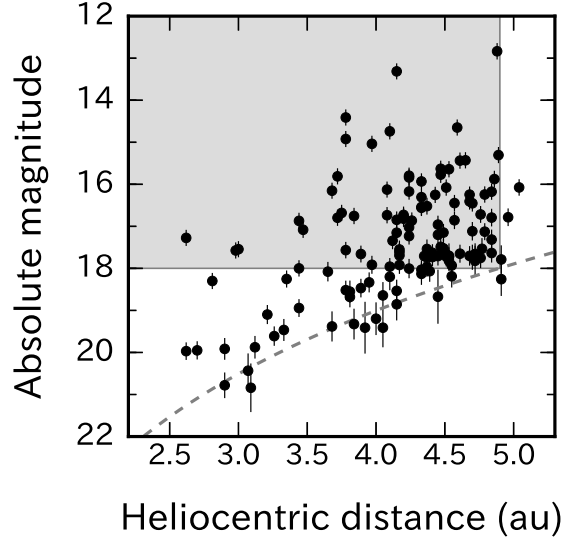}
\caption{
Heliocentric distance ($r$) vs. absolute magnitude ($H$) for the detected Hildas. 
The dashed line corresponds to apparent magnitude of 24.4~mag as the completion limit of this 
survey. 
The shaded area shows the range of our sample selection, $r$~$<$~4.9~au and $H$~$<$~18.0~mag. 
\label{fig03}
}
\end{figure}

Figure~\ref{fig03} shows the plot of the corrected heliocentric distance and absolute magnitude
of the detected Hildas.
The fainter (i.e., smaller) objects are clearly seen to have been detected at locations closer to 
the Sun due to the decrease in apparent brightness with increasing heliocentric/geocentric 
distances.
In addition, the airmass, sky condition, and vignetting cause nonuniformity of the detection
sensitivity in each frame and CCD, which needs to be corrected for accurate measurement of the
size distribution.
We evaluated the detection efficiency of all the frames on a CCD-by-CCD basis using synthetic 
moving objects imitating Hildas implanted into the images.
The completion limit of this survey was determined to be $m$~=~24.4~mag where the detection 
efficiencies achieved 50\% or more in all the data.

Then, we defined the outer edge of the sampling range of heliocentric distance as $r$~=~4.9~au,
where the limiting magnitude of $m$~=~24.4~mag corresponds to $H$~=~18.0~mag.
91 objects with $r$~$<$~4.9~au and $H$~$<$~18.0~mag were selected as an unbiased sample for use in 
exploring the size distribution.
The absolute magnitude of those objects ranges from 12.8~mag to 18.0~mag, corresponding to
$D$~$\sim$~1$-$14~km from Equation~(\ref{eq02}), most of which have body sizes less than 3~km in 
diameter.

\subsection{Size distribution} \label{sec03-2}

\begin{figure}
\figurenum{4}
\epsscale{1.00}
\plotone{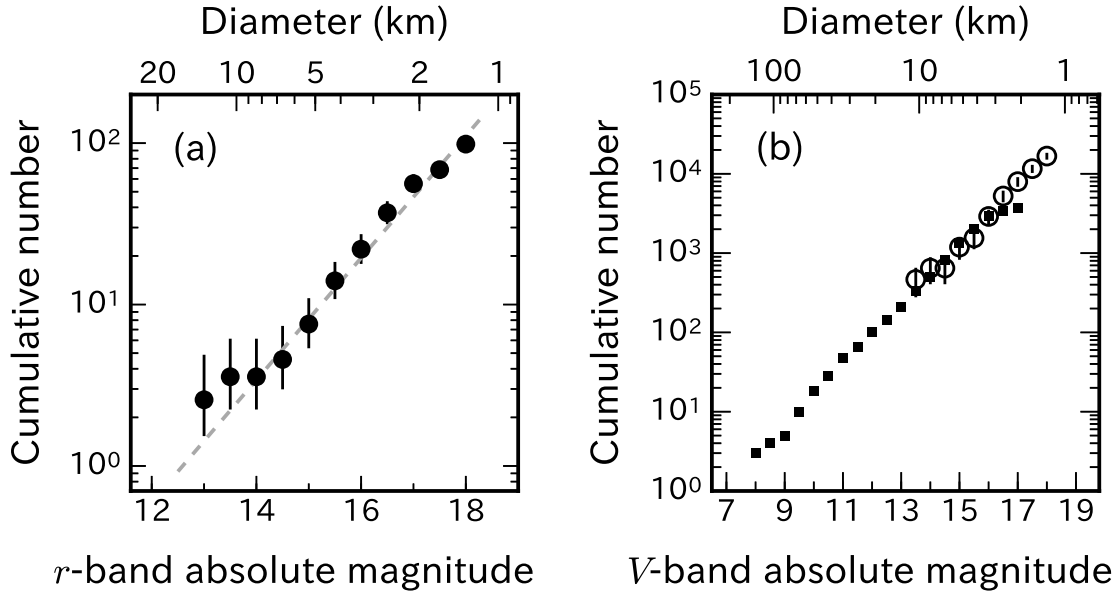}
\caption{
(a) The cumulative size distribution derived from the unbiased sample consisting of 91 Hildas. 
The cumulative numbers are corrected by the detection efficiency function.
The geometric albedo of each object is assumed to be 0.055. 
The dashed line shows the best-fit power law. 
(b) The cumulative size distributions of the Hilda population combined with the MPC catalog (squares) 
and this work (circles) scaled at $H_V$~=~16.0~mag which corresponds to $\sim$3.6~km in diameter.
\label{fig04}
}
\vspace{2em}
\end{figure}

We estimated the cumulative size distribution (CSD) of Hildas using the obtained sample consisting 
of 91 objects.
The cumulative number of these objects with absolute magnitude less than $H$, or $N ( < H )$, 
is corrected with the detection efficiency function.
Figure~\ref{fig04}(a) shows the CSD of our sample with error bars based on Poisson statistics of the 
cumulative number.
Although the object number is not sufficient for precisely determining the shape, the CSD seems to 
have an almost constant power-law slope over the diameter range of $\sim$1$-$10~km.

We approximated the differential $H$ distribution, $\Sigma ( H ) = dN(H)/dH$, of our sample
with a single-slope power law as
\begin{equation}
\Sigma ( H ) \, = \, 10^{\alpha ( H - H_0 )}, 
\label{eq03}
\end{equation}
where $\alpha$ is the power-law slope, and $H_0$ is given as $ \Sigma ( H_0 ) = 1$.
This distribution is also represented by cumulative number $N(>D)$, the number of objects larger 
than a diameter of $D$~km, as
\begin{equation}
N ( > D ) = N ( > 1 {\rm km} ) D^{-b},
\label{eq04}
\end{equation}
where $b = 5 \alpha$.
Power-law fitting was carried out by the maximum likelihood method \citep[e.g.,][]{Br04} as 
\citet{YT17} adopted.
Uncertainties in the fitted parameters are estimated from repeated fitting to synthetic object
samples generated from the actual objects based on the measurement errors. 

We obtained the best-fit power-law slope of $\alpha = 0.38 \pm 0.02$, corresponding to 
$b = 1.89^{+0.12}_{-0.11}$.
As seen in Figure~\ref{fig04}(a), the CSD of our sample is mostly coincident with the best-fit power 
law within Poisson statistics, suggesting that Hildas' size distribution follows a single-slope 
power law, at least in this size range.

Then, we combined Hildas' CSDs derived from our survey and the asteroid database published by Minor 
Planet Center (MPC)\footnote{\url{http://minorplanetcenter.net/db_search}} containing 4,075 Hildas 
as of the end of 2017, as illustrated in Figure~\ref{fig04}(b).
Since the MPC catalog shows the $V$-band magnitude, the $r$-band magnitude of our sample was 
required to be converted into the $V$-band magnitude.
The $m_V-m_r$ color of 0.25~mag was used for this conversion, which was derived from the average
color of the known Hildas with $m_r$~$<$~20.0~mag listed in the fourth release of the SDSS Moving 
Object Catalogue (MOC4)\footnote{\url{http://faculty.washington.edu/ivezic/sdssmoc/sdssmoc.html}}.
The cumulative numbers of our sample were scaled to that of the MPC catalog at $H_V$~=~16.0~mag.
The complied CSD appears no significant feature such as a break or roll over, but implies a 
single-slope power-law distribution over the full size range observed so far.
This may suggest that the Hilda population is in collisional equilibrium up to nearly the largest 
bodies.

Figure~\ref{fig04}(b) indicates that the Hilda population contains 
$\sim$(1.0~$\pm$~0.1)~$\times$~$10^4$ asteroids with $D$~$>$~2~km.
\citet{YT17} showed the L4 JT population larger than 2~km in diameter consisting of 
(3.5~$\pm$~0.2)~$\times$~$10^4$ asteroids.
By using the number ratio between L4 and L5 JTs presented by \citet{NY08}, this is converted into 
the total number of JTs with $D$~$>$~2~km reaching (5.4~$\pm$~0.5)~$\times$~$10^4$. 
Thus, the number ratio of Hildas to JTs is estimated be 0.18~$\pm$~0.03.

\section{Discussion} \label{sec04}

As mentioned in Section~\ref{sec02}, our survey detected a number of JTs in addition to Hildas.
\cite{YT17} have revealed that the size distribution of the JT population is well described by
a single-slope power law with $\alpha = 0.37 \pm 0.01$ or $b = 1.84 \pm 0.05$ from $\sim$2 to 10~km
in diameter.
The size distribution of Hildas measured with the same dataset by the present work can be also 
fitted by a single-slope power law with $\alpha = 0.38 \pm 0.02$ or $b = 1.89^{+0.12}_{-0.11}$.
The power-law slopes of Hildas and JTs in the common size range are coincident.

\begin{figure}
\figurenum{5}
\epsscale{0.50}
\plotone{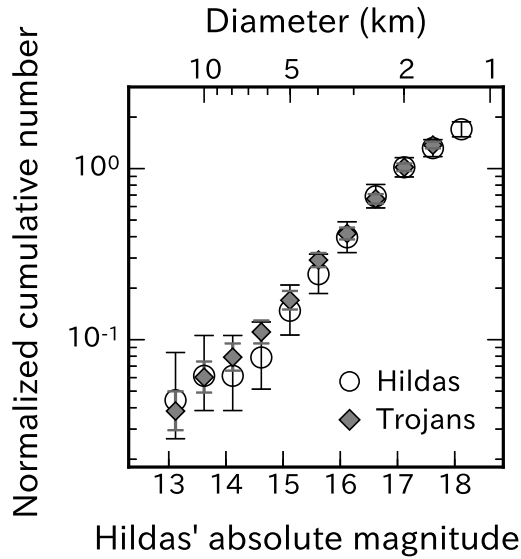}
\caption{
The cumulative size distributions derived from the Hilda (circles) and JT (diamonds) samples in 
our survey, both of which are normalized at 2~km in diameter.
The geometric albedos are assumed to be 0.055 for Hildas \citep{Gr12a} and 0.07 for JTs 
\citep{Gr12b}.
\label{fig05}
}
\end{figure}

For a more direct comparison, we normalized the CSDs of Hildas and JTs at $D$~=~2~km as shown 
in Figure~\ref{fig05}.
Note that the geometric albedos were assumed to be 0.055 for Hildas \citep{Gr12a} and 0.07 for 
JTs \citep{Gr12b}.
The two CSDs are in agreement within Poisson statistics of the cumulative numbers through the entire 
range.
We computed the two-sample Anderson-Darling (A-D) statistic \citep{Pt76} for the two CSDs to test 
the null hypothesis that these distributions are identical.
The value of the A-D statistic is 0.627, indicating that the null hypothesis cannot be rejected at 
even 60~\% significance level.
Considering these results, Hildas and JTs seem to have the same/similar size distribution shape
from 2~km or less to $\sim$10~km in diameter.

At diameter below a few tens km, these populations are not primordial but have been likely modified 
by collisional evolution \citep{dB07,GB08}.
Although the initial signature of the size distribution of km-sized objects has disappeared, we are 
still able to obtain alternative information about nature of the asteroid population from them. 
The size distribution of a mature collisionally-evolved population has a steady-state power law 
which is primarily determined by the impact strength law defined as the critical specific energy per 
unit target mass required for catastrophic disruption of the target \citep[e.g.,][]{Dr98,OG03}.
The strength law is directly linked with the internal physical properties which depend on the bulk 
composition and body structure.
The indistinguishable shapes of the size distributions indicates equivalent characteristics of 
body's interior between the two populations and thus is suggestive of a common formation environment.

Note that, as mentioned above, Hildas's size distribution exhibits a steady-state power law up to 
the largest body size (see Figure~\ref{fig04}b), which implies that the collisional evolution is 
dominated over the full size range.
That seems to be not the case with JTs because they have a clear transition at $D$~$\sim$~50~km
in power-law slope of the size distribution \citep{YT17}.
Therefore, a comparison of the size distributions between Hildas and JTs at large sizes is not 
useful to diagnose their similarity.

On the other hand, the size distribution of MBAs is obviously distinguishable from those of JTs and 
Hildas (\citealt{YT17}; Yoshida et al., submitted to P\&SS).
The power-law slope and local patterns (e.g., bump, dip, or knee) of a size distribution for a
small-body population in collisional equilibrium are primarily derived from properties of the 
critical specific energy for catastrophic disruption, so-called $Q^\ast_D$ law, following power laws 
with body size \citep{Dv02,OG03}.
The transition in the power-law slope of $Q^\ast_D$ between the gravity- and strength-scaled regimes 
induces a ``wave" pattern in the gravity-scaled portion of the size distribution as found in MBAs 
\citep[e.g.,][]{Dr93,Dv94,CB94}.
The size distributions of JTs and Hildas, however, show no significant wavy structure.
Meanwhile, the mean collisional velocities are similar among MBAs, Hildas, and JTs \citep{Dh98}.
These facts imply that Hildas/JTs have a different $Q^\ast_D$ (or shattering impact specific energy
$Q_S$ as \citet{dB07} pointed out) law, i.e., different bulk composition and/or internal structure, 
from MBAs, while the two populations appear to share the common properties.
Accordingly, we suggest that Hildas and JTs originate in the same population which does not 
correspond to MBAs.

This conclusion may lead us to infer that a major portion of Hildas and JTs were not formed in situ, 
but were transported from beyond Jupiter.
Our findings are qualitatively consistent with dynamical simulations for MBAs, Hildas, and JTs, 
based on the jumping Jupiter model.
\citet{Lv09} showed that a significant number of trans-Jovian planetesimals are captured into the 
Hilda/JT orbits through gravitational scattering induced by the migration of giant planets.
On another hand, \citet{RN15} showed very low survival probabilities of the in-situ populations and
minor contributions of the implanted populations from the outer main asteroid belt to the present 
Hilda/JT populations.
Further theoretical and observational studies are required for examining whether this model can 
account for the size distributions of small-body populations from Hildas to trans-Neptunian 
objects.

The comparable color distributions of non-family Hildas and JTs reported by \citet{WB17a} also
support this model.
Although \citet{Wn17} and \citet{DP18} pointed out a discrepancy in the visible/near-infrared 
spectral properties between Hildas and JTs, it may be due to their difference surface temperatures 
depending on heliocentric distance or to their different resurfacing frequency.

It is also worth noting that JTs' size distribution has a steeper power-law slope beyond $\sim$10~km 
in diameter than that in the km-size range \citep{YT17}, while Hildas' size distribution appears to 
maintain the constant slope over the full size range observed so far.
This discrepancy may indicate a difference in the progress degree of collisional evolution between
the two populations.
As mentioned in Section~\ref{sec03-2}, Hildas are almost wholly in collisional equilibrium.
In contrast, considering the analogical structure of MBAs' size distribution, the size distribution 
of large JTs with a steep slope seems to be predominantly unaffected by collisional fragmentation 
and to reflect that of the primordial population formed in the accretion process.
According to the simulation results in \citet{dB07}, the mean collisional lifetimes of L4 JT 
asteroids with $D$~$\gtrsim$~80~km are longer than the solar system age.
Therefore, JTs are considered to be less collisionally evolved than Hildas.
However, this suggestion is in conflict with the mean intrinsic collision probability for Hildas, 
which is lower than that for JTs \citep{Dh98}.
As discussed in \citet{WB17a}, the Hilda population in the early solar system might have suffered 
much more frequent collisions than the present and then been significantly depleted during the 
dynamical instability phase induced by planetary migration.
This could be caused by a huge number of impactors intruding from the main asteroid belt into the 
Hilda region.
At the time of the implantation of Hilda asteroids from between/beyond the giant planet regions,
MBAs may have consisted of more asteroids with excited orbits capable of reaching 4~au from the 
Sun than the present.

\acknowledgments
We thank an anonymous referee for helpful comments and suggestions.
This publication makes use of data collected by Subaru Telescope, which is operated by the National 
Astronomical Observatory of Japan (NAOJ). 
The HSC instrumentation and software were developed by NAOJ, the Kavli Institute for the Physics 
and Mathematics of the Universe (Kavli IPMU), the University of Tokyo, the High Energy Accelerator
Research Organization (KEK), the Academia Sinica Institute for Astronomy and Astrophysics in Taiwan 
(ASIAA), and Princeton University. 
We used \textit{hscPipe} (v~3.8.5) for data reduction.
This work was supported by JSPS KAKENHI Grant Numbers JP18K13607 (TT), JP15H03716, and 16K05546 (FY).


\end{document}